\documentclass{Interspeech2024}
\usepackage{url}
\usepackage{amsmath,graphicx}
\usepackage{hyperref}
\usepackage[dvipsnames]{xcolor}
\usepackage{booktabs}
\usepackage{amsfonts}
\usepackage{bm}
\usepackage{cleveref}
\usepackage{multirow}
\usepackage{lipsum}
\usepackage{makecell}
\usepackage{caption}

\interspeechcameraready 
\newcommand\blfootnote[1]{%
  \begingroup
  \renewcommand\thefootnote{}\footnote{#1}%
  \addtocounter{footnote}{-1}%
  \endgroup
}

\title{On the Utility of Speech and Audio Foundation Models for Marmoset Call Analysis}

\name[affiliation={1,2}]{Eklavya}{Sarkar}
\name[affiliation={1}]{Mathew}{Magimai.-Doss}

\address{
  $^1$Idiap Research Institute, Switzerland\\
  $^2$Ecole polytechnique fédérale de Lausanne, Switzerland
  }
\email{\{eklavya.sarkar, mathew\}@idiap.ch}


\begin{document}


\maketitle

\begin{abstract}
    Marmoset monkeys encode vital information in their calls and serve as a surrogate model for neuro-biologists to understand the evolutionary origins of human vocal communication. Traditionally analyzed with signal processing-based features, recent approaches have utilized self-supervised models pre-trained on human speech for feature extraction, capitalizing on their ability to learn a signal's intrinsic structure independently of its acoustic domain. However, the utility of such foundation models remains unclear for marmoset call analysis in terms of multi-class classification, bandwidth, and pre-training domain. This study assesses feature representations derived from speech and general audio domains, across pre-training bandwidths of 4, 8, and 16 kHz for marmoset call-type and caller classification tasks. Results show that models with higher bandwidth improve performance, and pre-training on speech or general audio yields comparable results, improving over a spectral baseline.
\end{abstract}
bioacoustics, call-type and caller classification, speech and audio, bandwidth.

\section{Marmoset Vocalizations}
\begin{figure*}[ht]
  \centering
  \centerline{\includegraphics[width=\textwidth]{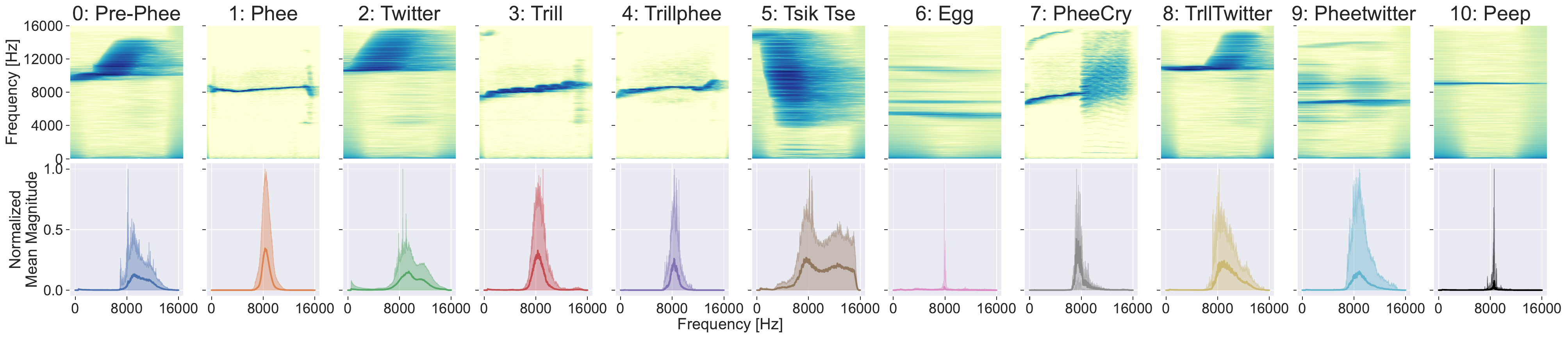}}
  \caption{Marmoset vocalizations with a 16 kHz bandwidth. {\normalfont{Top:}} Spectrograms of a {\normalfont{single}} call-type vocalization. {\normalfont{Bottom:}} The mean spectrum for {\normalfont{all}} vocalizations per call-type across the dataset, normalized. Shaded areas indicate $\pm$ 1 std from the mean spectrum.}
  \label{fig:vocs}
\end{figure*}
\blfootnote{\href{https://github.com/idiap/speech-utility-bioacoustics}{\url{github.com/idiap/speech-utility-bioacoustics}}}
Non-human vocal communication, such as bioacoustics, i.e. the study of animal vocalizations, is rapidly advancing through the advent of machine learning and the correlated progress in human speech processing \cite{bioacoustics_roadmap}. Common marmosets (\textit{Callithrix jacchus}) are of particular interest due to their highly vocal nature, acoustically diverse call repertoire, and acute auditory capabilities. Their extensive vocalizations are rooted in a complex social system, and are thus able to encode a range of information, such as group affiliation, sex \cite{Norcross1993}, population, dialect \cite{Zurcher2017}, and even individual caller identity \cite{Jones1993, Phaniraj2023}, over a number of social and emotional states \cite{Epple1968, seyfarth2003}. Their remarkable vocal adaptability also allows them to modify the duration \cite{brumm2004}, intensity \cite{Eliades2012}, complexity \cite{Pomberger2018}, or timing \cite{Roy2011} of their calls. These vocal characteristics align them closely with human speech properties, such as care-giving to infants, turn-taking \cite{takahashi_2016}, and categorical perception of sounds \cite{OsmanskiWang2023}, and make them into a well-suited surrogate model for understanding the vocal communication of non-human primates among biologists~\cite{marmoset_biolgy_2014} and neuroscientists~\cite{brain_minds}.

In the literature, the automatic analysis of marmoset vocalizations, i.e. call-type, caller identity, or sex classification, has been conducted by leveraging signal processing features alongside traditional machine learning classifiers. Early work demonstrated that k-NN, SVM, and optimal path forest classifiers achieved notable success over multilayer perceptrons (MLPs), Adaboost, and logistic regression, especially with small, specific datasets~\cite{lpc_ML_marmoset_2016}. Research exploring a variety of audio and spectral feature representations, such as signal energy, zero crossing rate, spectral rolloff, and MFCCs, indicated that integrating different feature could enhance the system's performance on synthetically augmented vocal datasets~\cite{wisler16_interspeech}. Recent studies have also explored leveraging deep learning based techniques. Using convolutional neural networks to process spectrograms for simultaneous vocalization detection, call-type classification, and caller identification was found to outperform separate models for each task \cite{Oikarinen_jasa_2019}. Statistics of log-mel filter-bank energies used as input for recurrent neural networks (RNNs) were shown to improve the detection and classification of calls over SVM or MLPs \cite{cas}. Self-supervised learning (SSL) frameworks, which create surrogate labels from the data, were used with the aim of leveraging the large quantities of unlabeled data for birdsong detection~\cite{cola_paper} and bioacoustic event detection~\cite{bioacoustic_event}. 

A novel study demonstrated that neural representations derived from models pre-trained on human speech through SSL could distinguish individual marmoset caller identities \cite{sarkar23_interspeech}. The authors argued that SSLs only learn the intrinsic structure of the unlabeled input signal, typically through a masking-based pre-text training task, to capture essential information independently of any domain-specific knowledge, such as human speech production, and thus can be cross-transferred across different acoustic domains, such as bioacoustics. Building on these findings, our paper investigates the utility and limitations of such pre-trained foundation models for the purpose of marmoset call analysis, with a focus on the following key points:
\begin{enumerate}
\item \textbf{Classification}: We investigate whether such models can be effectively leveraged for marmoset call analysis tasks, namely call-type and caller classification, which, to the best of our knowledge, has not yet been demonstrated. Additionally, while \cite{sarkar23_interspeech} focused solely on caller detection in a binary framework, we extend the scope to a multi-class approach.
\item \textbf{Bandwidth}: Given that these models are typically pre-trained at a bandwidth of 8 kHz, we address their mismatch with the biological vocalization and auditory range of marmosets, predominantly concentrated in the 5--10 kHz spectral region~\cite{OSMANSKI20161}, and thus evaluate their capability to accurately represent marmoset calls. By examining models pre-trained across varying bandwidths, we aim to evaluate their effectiveness in adequately representing marmoset calls, and seek to clarify how model bandwidth influences their classification.
\item \textbf{Pre-training domain}: It remains unclear how models pre-trained on human speech compare to trained on other acoustic domains for accurately capturing marmoset call characteristics. We examine representations produced by different pre-training sources, such as human speech and general audio, across supervised and self-supervised learning frameworks, against a spectral baseline to identify the most suitable pre-training source for cross-domain bioacoustic signal analysis.
\end{enumerate}
The rest of the paper is organized as follows: \Cref{sec:metho} gives the study's methodology, \cref{sec:sim_anal} \& \ref{sec:class_anal} present a call similarity and classification analysis. \Cref{sec:concl} finally concludes the paper.

\section{Methodology}\label{sec:metho}

\subsection{Dataset and Tasks}
For our study, we used the InfantMarmosetsVox (IMV) dataset \cite{sarkar23_interspeech}, which contains $72,921$ labelled marmoset vocalization segments (totalling to 464 minutes), sampled at $44.1$ kHz, across ten marmoset individuals and contains eleven marmoset call-types. \Cref{table:dataset_stats} presents the data distribution in function of the call-types and callers. For our experiments, we divide the dataset into a \textit{Train}, \textit{Val}, and \textit{Test} sets, following a random 70:20:10 split. We denote call-type and caller identity multi-class classification as CTID and CLID respectively.
\begin{table}[ht]
\centering
\caption{InfantMarmosetsVox dataset statistics.}
\begin{tabular}{lll}
\toprule
\textbf{ID} & \textbf{Call-type} & \textbf{Count} \\
\midrule
0  & Peep (pre-phee) & 1283  \\
1  & Phee            & 27976 \\
2  & Twitter         & 36582 \\
3  & Trill           & 1408  \\
4  & Trillphee       & 728   \\
5  & Tsik Tse        & 686   \\
6  & Egg             & 1676  \\
7  & Pheecry (cry)   & 23    \\
8  & TrllTwitter     & 293   \\
9  & Pheetwitter     & 2064  \\
10 & Peep            & 202   \\
\midrule
& \textbf{Total} & \textbf{72921} \\
\bottomrule
\end{tabular}
\quad
\begin{tabular}{cl}
\toprule
\textbf{Caller ID} & \textbf{Count} \\
\midrule
0 & 15521 \\
1 &  8648 \\
2 & 13827 \\
3 &  5838 \\
4 &  5654 \\
5 &  3522 \\
6 &  4389 \\
7 &  2681 \\
8 &  6387 \\
9 &  6454 \\
- & - \\
\midrule
\textbf{Total} & \textbf{72921} \\
\bottomrule
\end{tabular}
\label{table:dataset_stats}
\end{table}

\Cref{fig:vocs} gives the visualizations of all call-types as well the density distribution of the spectrums across the entire dataset. Frequencies below 500 Hz are nullified purely for visualization to eliminate any low-frequency noise. We can observe that information starts at around 7-8 kHz for most calls in this dataset.

\subsection{Models and Feature Representations}
For our study, we select four distinct frameworks for feature representations $\mathcal{F}$: hand-crafted (HC) features derived through signal processing techniques, neural representations obtained via self-supervised learning (SSL), pre-trained on either human speech or general audio, and features generated through supervised learning (SL) models pre-trained on general audio. These frameworks are summarized in \cref{table:models}. We extract the features from these frameworks by giving the marmoset calls as input.
\begin{table}[ht]
\centering
\caption{\# Parameters $P$ and feature dimension $D$ of selected models, pre-trained on AudioSet (AS) or LibriSpeech (LS).}
\begin{tabular}{rclll}
\toprule
\bm{$\mathcal{F}$} & \textbf{Corpus} & $\bm{P}$ & $\bm{D}$ & \textbf{Type} \\
\midrule
C22 \cite{lubba2019catch22}     & -      & -     & 24    & HC \\
WavLM \cite{wavlm}              & LS     & 94.38M & 1536  & SSL\\
BYOL \cite{niizumi2021byol-a}   & AS     & 5.32M  & 2048  & SSL\\
PANN \cite{PANN}                & AS     & 8.08M  & 2048  & SL \\
\bottomrule
\end{tabular}
\label{table:models}
\end{table}

\textbf{Hand-crafted}: The Highly Comparable Time-Series Analysis (HCTSA) framework, used for interpreting diverse time series data, extracts 7700 features through signal processing methods, such as LPC \cite{hctsa_2013}. It has been applied to diverse tasks such as birdsong discrimination~\cite{birdsong_hctsa}, ecosystem monitoring~\cite{sethi2020automated}, and marmoset caller identification~\cite{Phaniraj2023}. Despite its broad applicability, HCTSA's computational demands and feature redundancy are significant limitations. The CAnonical Time-series CHaracteristics (Catch22/C22), a steamlined subset of HCTSA, provides high performance with minimal redundancy across numerous classification problems~\cite{lubba2019catch22}. We extend this feature set to a final dimension of $D = 24$ by appending the first and second order statistics, and use it as our spectral baseline.

\begin{figure*}[ht]
  \centering
  \includegraphics[width=\linewidth]{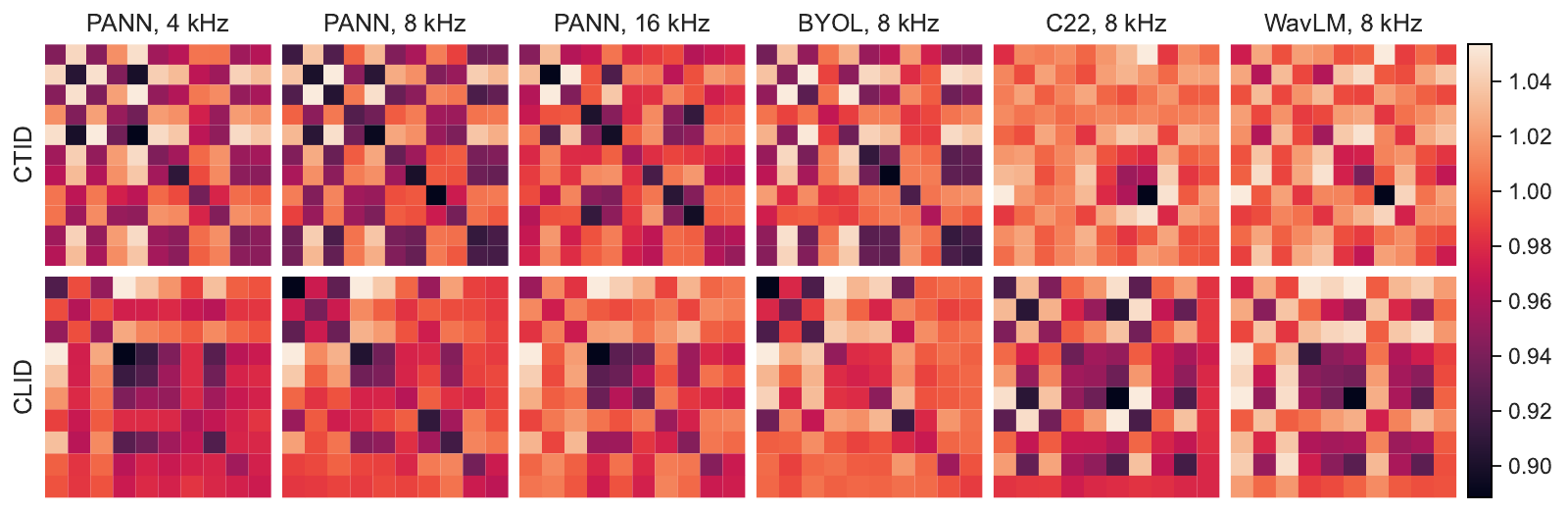}
  \caption{Pairwise mean cosine distances matrices for features $\mathcal{F}$ at different bandwidths for call-types (CTID) and callers (CLID). Diagonal entries represent intra-class distances, and off-diagonal the inter-class. Darker regions indicate higher similarity.}
  \label{fig:dists}
\end{figure*}

\textbf{SSL pre-trained on human speech}: Following the approach in~\cite{sarkar23_interspeech}, we use feature representations from SSL models trained on human speech, extending it to both call-type and caller identity classification. We select the WavLM base model, pre-trained on the 960-hour LibriSpeech dataset, based on its effectiveness in marmoset call detection as well as its versatility in speech processing tasks as demonstrated in the SUPERB challenge~\cite{yang21c_interspeech}. For each layer, feature representations of length $768$ are extracted for each frame. Then, they are transformed into fixed-length utterance-level representations by computing and aggregating first and second order statistics across the frame-axis, resulting in a final representation of length $D=1536$.

\textbf{SL pre-trained on general audio}: Expanding marmoset call analysis literature, we utilize embeddings from models pre-trained on the AudioSet (AS) dataset, which includes audio event classes such as environmental sounds, musical instruments, and human and animal vocalizations. Specifically, we choose the \textit{AudioNTT2020} model from the BYOL-A architecture \cite{niizumi2021byol-a}, extracting embeddings from its final fully connected layer of length $D=2048$. Inputs are processed into log-mel spectrograms, adhering to the spectral parameters detailed in the original study, i.e. a 8 kHz bandwidth, 64 ms window size, 10 ms hop size, and 64 mel bins spanning from 60 to 7800 Hz.

\textbf{SL pre-trained on general audio}: We further investigate feature extraction from large-scale networks pre-trained for general audio pattern recognition. The \textit{CNN14} model from the \textit{PANN} network \cite{PANN} is chosen, with pre-trained weights applied at three different bandwidths: 4, 8, and 16 kHz. This model employs a balanced sampling strategy across AudioSet's sound classes and also processes input vocalizations into spectrograms to extract log-mel filterbanks. For a bandwidth of 16 kHz, window and hop sizes are set to 1024 and 320 samples, respectively, and proportionally halved for 8 and 4 kHz. The model utilizes 64 mel bands, spanning from 50 Hz and to the Nyquist frequency. Embeddings of length $D=2048$ are extracted from the linear layer preceding the final classification layer.


\section{Call Similarity Analysis}\label{sec:sim_anal}
This section presents a pairwise similarity analysis of the selected features on the \textit{Train} set to identify any discernible patterns or correlations for given the vocalizations. Specifically, we investigate how variations in the bandwidth of the pre-trained models affect the similarity distribution of intra-class embeddings, and examine any distinctions between models pre-trained on speech against general audio. To compare the features, which are high-dimensional vectors, we use the cosine distance defined as $\text{sim}(\bm{x_1}, \bm{x_2})= 1 - (\bm{x_1} \cdot \bm{x_2} \mathbin{/} \lVert \bm{x_1} \rVert \cdot \lVert \bm{x_2} \rVert)$, bounded in $[0, 2]$. Two features are identical when their cosine distance is 0, orthogonal at 1, and opposite at 2. For WavLM, we select the first layer, and only use the first half of the extracted features, corresponding to the mean values averaged frame-wise.
\begin{figure}[ht]
  \centering
  \includegraphics[width=\linewidth]{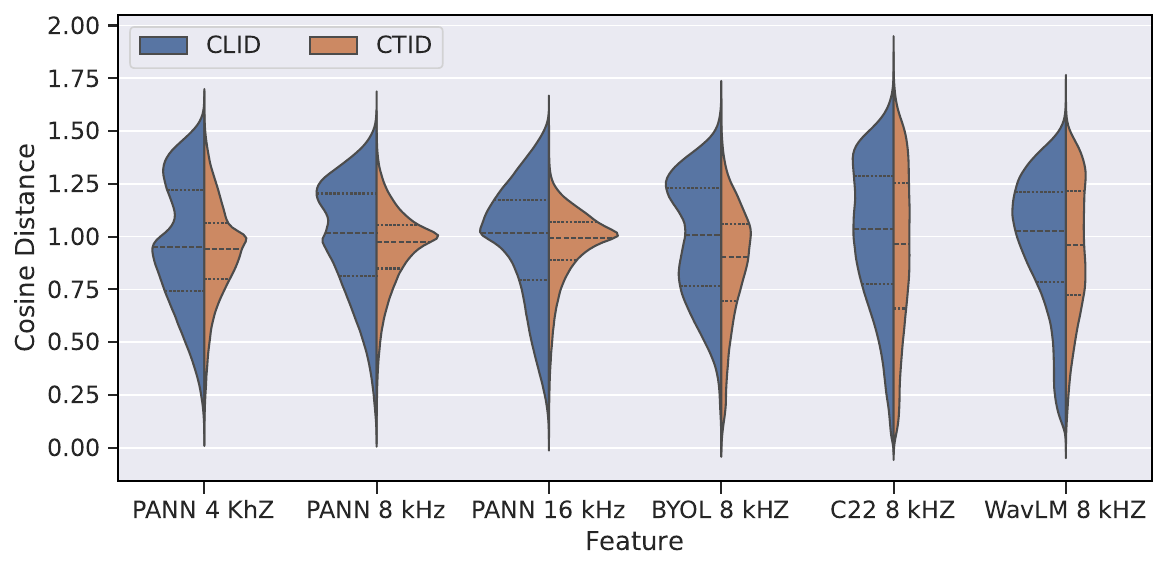}
  \caption{Distribution of pairwise cosine distances.}
  \label{fig:violin_dists}
\end{figure}

\Cref{fig:violin_dists} presents the overall distribution of pairwise distances. The distrtibutions are overlapping, centering around a median distance of 1 for all representations, suggesting a lack of clear correlation or similarity within the embeddings generated. \Cref{fig:dists} further delineates the distributions into distance matrices for each feature set, where diagonal and off-diagonal entries correspond to intra-class and inter-class distances respectively. In an ideal scenario, embeddings from the same call-type or caller would exhibit closer distances, where as embeddings from different classes would have a higher dissimilarity. 

We can observe that the models pre-trained on general audio datasets (BYOL and PANN) yield more distinct peaks and diagonals, on figures \ref{fig:violin_dists} and \ref{fig:dists} respectively, compared to those pre-trained on human speech (WavLM) or the handcrafted baseline (Catch22). This distinction is more pronounced for call-types than for caller identification. This is expected, given that the call-types are spread across caller classes (a caller produces different calls, while a call can come from any caller). Although these patterns indicate some level of class-specific clustering, the distribution of distances largely show that the features are highly orthogonal. The similarity analysis thus indicates minimal feature correlation, and suggests that classifying these vocalizations with a simple linear classifier would be challenging, as there is no clear linear separability between the classes.
\begin{figure*}[ht!]
  \centering
  \includegraphics[width=\linewidth]{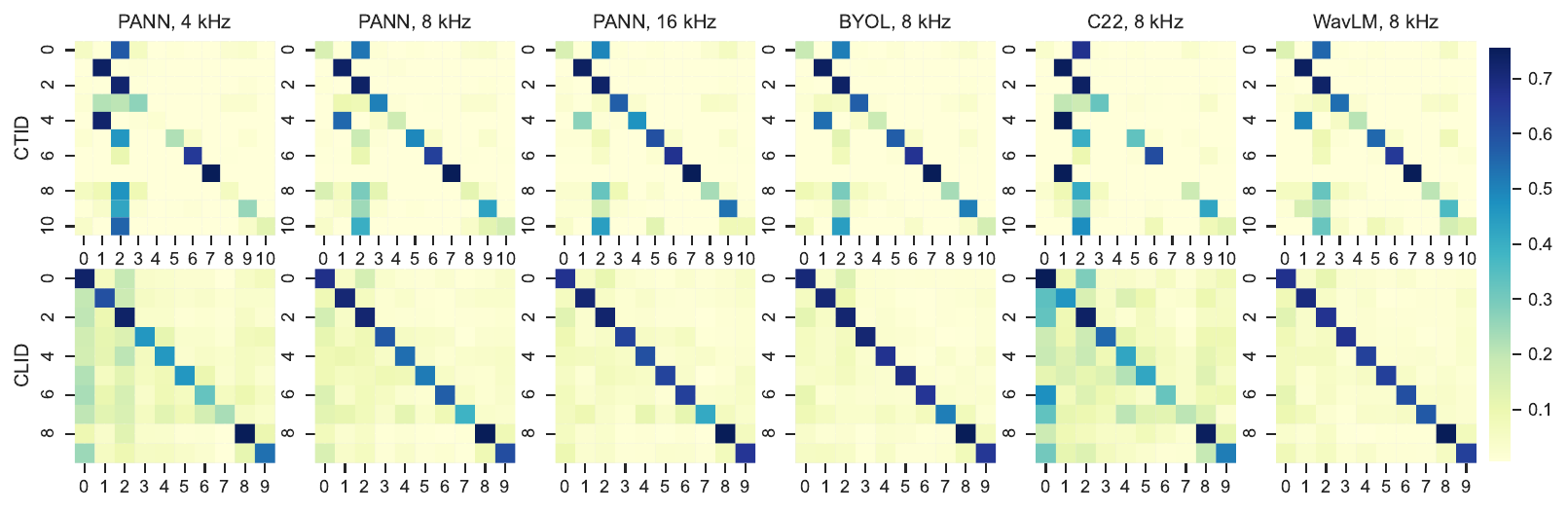}
  \caption{Normalized confusion matrices with row indices representing true class labels. Darker diagonals signify higher performance.}
  \label{fig:cms}
\end{figure*}

\section{Classification Analysis}\label{sec:class_anal}
Based on the insights of our similarity analysis, we aim to evaluate the saliency of the extracted representations, and proceed to classify them using a simple, non-linear MLP, for the multi-class classification tasks. We implement three blocks of [Linear, LayerNorm, ReLU] layers, with 128, 64, and 32 number of hidden units respectively, followed by a final linear layer to obtain the posterior probabilities. To evaluate the performance we used Unweighted Average Recall (UAR) as the metric to account for any class imbalance. To obtain robust results, we employ the grid search methodology with \textit{Val} UAR score as the optimization criterion. We train the classifier for 30 epochs with cross-entropy loss, and search for the optimal hyperparameters values of $\eta$ and batch-size across $2^{\lbrack5\text{\textendash}9\rbrack}$ and [1e-3, 1e-4] respectively for each feature--task permutation on \textit{Train} and \textit{Val}. The optimization consists of Adam and a $\eta$-scheduler of factor $0.1$ and patience of $10$ epochs. Lastly, for WavLM, we classify each of the encoder layers [0\textendash13] to identify the optimal layer.

\begin{figure}[ht]
  \centering
  \includegraphics[width=\linewidth]{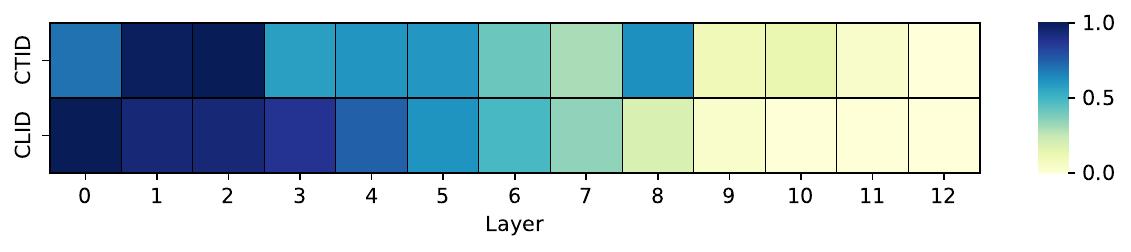}
  \caption{Layer-wise UAR scores of WavLM features, normalized per task. Darker regions indicate a higher performance.}
  \label{fig:wavlm_layers}
\end{figure}

\Cref{fig:wavlm_layers} presents the layer-wise scores for WavLM, normalized per task to a [0, 1] range. We can observe that the lower layers are clearly much more salient representations for both tasks compared to higher layers. Based on these results, we use the best individual WavLM layers for our two tasks. 

\Cref{table:uar_results}a) summarizes the classification results of the different feature sets at an 8 kHz bandwidth (BW). Random performance is given as 100 over the number of classes. Notably, BYOL features outperform the other features, for both CTID and CLID, despite having fewer parameters than WavLM and PANN, while C22 proves to be the overall weakest representation. WavLM shows the highest difference in performance across tasks. Meanwhile, \cref{table:uar_results}b) highlights the impact of pre-training bandwidth for salient representations on PANN features. The results clearly show that the bandwidth size correlates directly with the performance, increasing monotonically. Particularly, PANN features at 16 kHz achieve the highest performance across all features and BWs for CTID. BYOL embeddings at 8 kHz notably outperform PANN at 16 kHz for CLID. The best scores for both tasks are also closely matched in value.

\begin{table}[!htb]
\centering
\caption{UAR scores [\%] on \textit{Test} for pre-trained features $\mathcal{F}$. WavLM's best layer's score is given.}
\begin{tabular}{ccccc}
\toprule
Section & \bm{$\mathcal{F}$} & \textbf{BW} & \textbf{CTID} &  \textbf{CLID} \\
\midrule
\multirow{5}{*}{(a)} & Random & - & 9.09 & 10 \\
& C22   & 8 & 41.96  & 35.62 \\
& WavLM & 8 & 59.99 & 67.47 \\
& BYOL  & 8 & \textbf{63.64} & \textbf{68.30} \\
& PANN  & 8 & 58.54 & 56.02 \\
\midrule
\midrule
\multirow{3}{*}{(b)} & PANN  & 4  & 46.27 & 41.10 \\
& PANN  & 8  & 58.54 & 56.02 \\
& PANN  & 16 & \textbf{69.09} & \textbf{65.39} \\
\bottomrule
\end{tabular}
\label{table:uar_results}
\end{table}

\Cref{fig:cms} shows the classifier's performance through confusion matrices. We can again clearly observe the monotonic improvement in CTID classification performance for PANN features as the bandwidth increases. We also notice a prevalent trend of false positives for call-type ID 2 (Twitter) across all feature sets, especially against IDs 0, 8, and 10, attributable to its high occurrence in the dataset and broad spectral range \cite{Pistorio2006, cmu_marmoset_journal}. The CLID results contain distinctly fewer misclassifications, which aligns with expectations since the call-types are spread among the different callers classes. The exception is C22, which yields the weakest performance. Caller classes with higher data volumes (IDs 0 and 2) perform better compared to the others. Finally, a clear improvement in performance correlated with bandwidth is seen for PANN features, as with CTID.

\section{Summary and Conclusion}\label{sec:concl}
This paper investigated the utility and limitations of foundations models, pre-trained on human speech or general audio, which have not been demonstrated for marmoset call-type and caller identity multi-class classification. To that end, we conducted and validated two studies across three lines of investigation.

First we conducted a call similarity analysis, which revealed that the features extracted from these models lacked linear separability within or across classes. Then, we conducted a classification study which demonstrated that a non-linear classifier can still achieve substantial performance, and highlighted that a larger bandwidth directly correlates with improved performance. Classification of call-types also appeared to be more sensitive to bandwidth changes than caller identities. Additionally, the pre-training domain of speech and general audio showed comparable performances, with a distinct improvement over handcrafted features. Finally, we obtained close best performance for both call-type and caller classification tasks.

In conclusion, our findings underscore the potential of leveraging pre-trained foundation models for bioacoustic signals, particularly when the model's bandwidth aligns with the biological auditory and vocal range of the studied species. Future collaborative work with biologists and linguistics researchers could explore the biological implications of these results, especially in understanding the evolutionary aspects of marmoset vocal behaviour and their perceptual processing, to bridge the gap between computational models and biological insights in non-human vocal communication research.

\section{Acknowledgements}
This work was funded by Swiss National Science Foundation's NCCR Evolving Language project (grant no. 51NF40\_180888).

\bibliographystyle{IEEEtran}
\bibliography{mybib}

\end{document}